# SIMULATION OF RETURNING ELECTRONS FROM A KLYSTRON COLLECTOR

Z. Fang, The Graduate University For Advanced Studies
S. Fukuda, S. Yamaguchi, And S. Anami, High Energy Accelerator Research Organization (KEK)
1-1, Oho, Tsukuba-Shi, Ibaraki, 305-0801, Japan

*Abstract*

Spurious oscillations in klystrons due to returning electrons from the collector into the drift tube were observed and studied at KEK. Simulations of returning electrons using EGS4 Monte Carlo method have been performed. And the oscillation conditions are described in this paper.

## 1 INTRODUCTION

A high-power 324MHz klystron (3MW output power, 650μs pulse width, and 110kV beam voltage) is being developed at KEK as a microwave source for the 200 MeV Linac of the KEK/JEARI Joint Project for the high-intensity proton accelerator facility. However during high-voltage processing of the klystron tube #1, strong spurious oscillations were observed when there was no input power. These oscillations were caused by the returning electrons from the collector into the drift tube of the klystron. As the returning electrons modulated by the gap voltage of the output cavity will induce a gap voltage in the input cavity, an input cavity voltage is possibly regenerated by the returning electrons to cause the oscillations.

In order to examine the conditions of these oscillations, a simulation of returning electrons was conducted using EGS4[1]. The coefficients and energy distributions of the returning electrons were investigated at a range of beam voltages and with a variety of collector shapes, in order to examine the oscillation conditions thoroughly.

## 2 OSCILLATION PHENOMENA AND OSCILLATION SOURCE

During the high-voltage processing of the 324MHz klystron tube #1, unexpected oscillations were observed from both the output and input cavities when there was no driving input power. These oscillations occurred when the beam voltage was either 63~71kV or larger than 90kV, and had a frequency close to 324MHz.

When a magnetic field was applied at the collector region to deflect the electron beam, the oscillations stopped. This indicates that the source of these oscillations was the collector region. However, as there was no resonance of frequency close to 324MHz corresponding to the dimensions of the collector, the oscillations were assumed to be the result of returning electrons.

The collector size was accordingly changed to evaluate its effect on the returning electrons. A collector with a larger radius and longer in length was used in the tube #1A, where the results of the experiment indicated that the oscillations disappeared for the low beam voltage region and started from 95kV. When the length of the collector was increased further in the tube #2, again there were no oscillations at the low beam voltage region, and these did not occur until the beam voltage level was even higher. The shapes of the collectors are shown in Fig.1 and the oscillation regions for the three tubes are summarized in table 1.

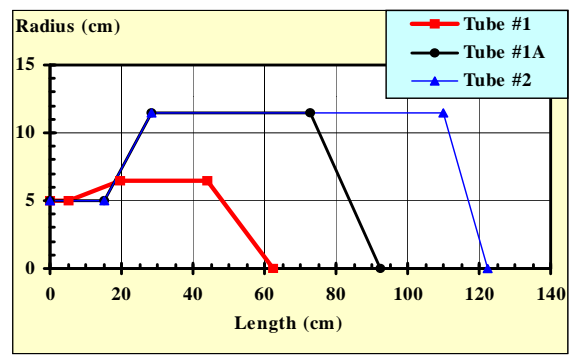

Figure 1: Collector shapes of tubes #1, #1A, and #2.

Table 1: Collector shapes and experiment results of the beam voltage regions of the oscillations.

| Tube | Collector radius(cm) | Collector length(cm) | Beam voltage regions of the oscillations (kV) |
|---|---|---|---|
| #1 | 6.5 | 62.4 | 63<V<71,V>90 |
| #1A | 11.5 | 92.4 | V>95 |
| #2 | 11.5 | 122.4 | V>104 or larger |

# 3 OSCILLATION STUDY

## 3.1 Oscillation Mechanism

In the klystron, after the electrons of the main beam bombard the surface of the collector, back-scattered electrons are produced due to an inelastic process with atomic electrons, and some of these back-scattered electrons return into the drift tube of the klystron. As these returning electrons are modulated by the gap voltage of the output cavity(**Vo**), they induce an rf voltage in the input cavity(**Vf**). In this way, the returning electrons and the main beam of the klystron form a feedback loop(Fig.2) and it's possible to cause spurious oscillations.

In Fig.2, **A** is the voltage gain induced by the main beam of the klystron; **A** = **Vo**/**Vd**, where **Vd** is the input cavity voltage modulating the main beam. **β** is the voltage gain induced by the returning electron current; **β** = **Vf**/**Vo**. |**β**| is proportional to the returning electron current as the onset of the oscillations is in the small-signal linear region.

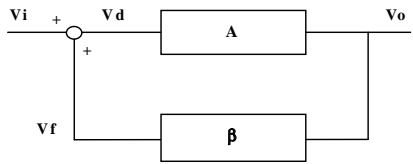

Figure 2: Block diagram of the feedback loop of the klystron due to the returning electrons.

## 3.2 Simulations of Returning Electrons

The EGS4(Electron Gamma Shower) Monte Carlo method was applied to the simulation of back-scattered and returning electrons in a klystron collector with an external magnetic field[2]. The trajectories for the incident electron beam were calculated initially with space-charge forces, relativistic effects, self-field and external magnetic fields. The cathode radius was 4.5cm and the magnetic field on the cathode surface was 130Gauss. The beam radius was 3.5cm at the entrance of collectors and the focusing magnetic field on the Z axis in the collector region is shown in Fig.3. The conditions at the start of the EGS4 Monte Carlo simulation were based on the above calculations for the incident electron positions, and their velocities and energies when bombarding the surface of the collector. The trajectories of the incident electron beam, and the back-scattered and returning electrons are shown in Fig.4 for tubes #1, #1A, and #2.

Calculations of the coefficients and distributions of the Z-component of the energy($E_z$) normalized by the beam voltage($E_o$) for the returning electrons revealed that the coefficients and the peaks of the energy distributions for $E_z/E_o$ were also identical under different beam voltages for a given collector.

Simulations of the returning electrons were carried out with a variety of collector shapes. When the size of the collector was enlarged, the amount of returning electrons decreased. The returning coefficients were 0.68%, 0.17%, and 0.14% for the collectors of tubes #1, #1A, and #2, respectively. The energy distributions for $E_z/E_o$ are shown in Fig.5. The shapes of the distributions differ slightly between the different collectors.

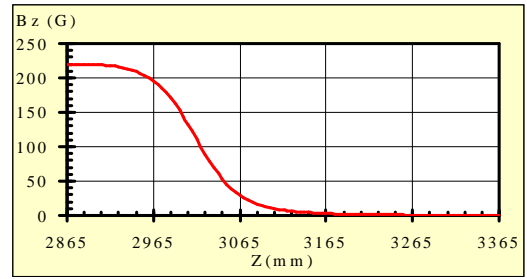

Figure 3: Magnetic field on the Z axis in the collector region.

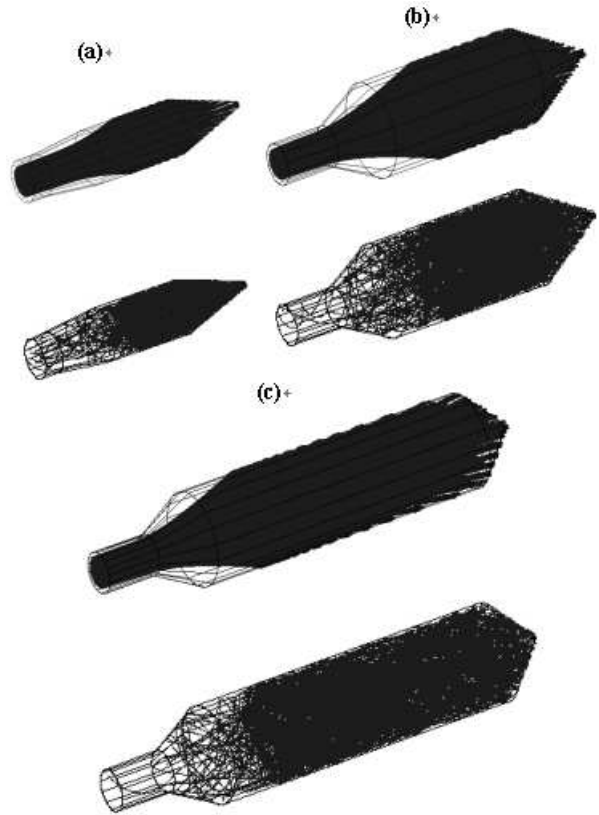

Figure 4: Trajectories of the incident electron beam, and the back-scattered and returning electrons in the collector of tubes #1(a), #1A(b), and #2(c).

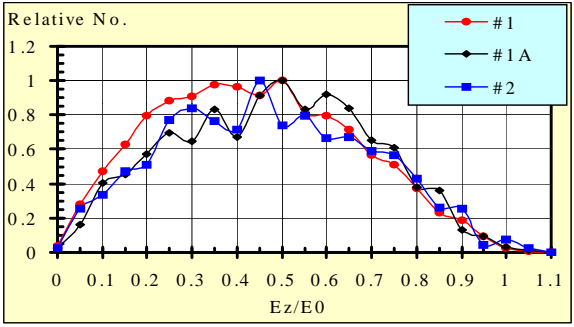

Figure 5: Energy distribution of the returning electrons of tubes #1, #1A, and #2.

### 3.3 Oscillation Conditions

Returning electrons modulated by the output cavity voltage will induce an rf voltage in the input cavity. If the phase of this induced voltage is the same as the phase of the voltage modulating the main beam, and the amplitude is larger than the latter, then an input cavity voltage is regenerated by the returning electrons, and spurious oscillations will occur.

In other words, the oscillation conditions are:

$$|A\beta| > 1 \quad (1)$$
$$\arg(A\beta) = \arg(A) + \arg(\beta) = 2n\pi, \quad n = 0, 1, 2\ldots \quad (2)$$

where the complex variables $A$ and $\beta$ are the voltage gains induced by the main beam and returning electrons respectively.

$A = |V_o/V_d| \angle -\theta_0$
$\beta = |V_f/V_o| \angle -\theta_0'$

where $\theta_0$ and $\theta_0'$ are the delayed angles corresponding to the DC transit angles, which are a function of frequency and beam voltage.

$A$ can be obtained using the klystron simulation code JPNDISK, the one-dimensional disk-model code.

$\beta$ can be calculated from the rf current of the returning electrons. As the returning electron current is very small, and the onset of the spurious oscillations is in the small-signal-linear region, it is possible to apply the ballistic theory. As the contribution from the output cavity voltage is larger than the others, we just consider the output and input cavities.

It is necessary to modify the formula for the rf current in the ballistic theory for the returning electrons, to take into consideration the energy distributions, $\eta(x)$, of the returning electrons, which is a polynomial function fitted to the distribution shape for each of the collectors, as shown in Fig.5.

$$I_{rf} = \frac{2I_b \int \eta(x) J_1(X') \cos(\omega t_2 - \theta') dx}{\int \eta(x) dx}$$

where $x = E_z/E_o$, $I_b$ is the total current for the returning electrons, $J_1(X')$ is the Bessel function, and $X'$ is the bunching parameter for the returning electrons.

Fig.6 shows the curve of $A(\omega)\beta(\omega)$ for tube #1 as the frequency changes from 322 to 326MHz under beam voltages of 65kV, 70kV, and 75kV. We can see that, between 65 and 70kV, some frequency components exist that satisfy the oscillation conditions, $|A(\omega)\beta(\omega)| > 1$ and $\arg[(A(\omega)\beta(\omega)] = 2n\pi$.

The oscillation regions of tube #1 obtained from the calculations were 64~70kV and larger than 79kV. For tubes #1A and #2, the calculations indicate that the oscillation regions of beam voltage larger than 100kV and 105kV, respectively, due to decreases in the current for the returning electron. These calculations are close to values obtained for the oscillation regions in the experiments.

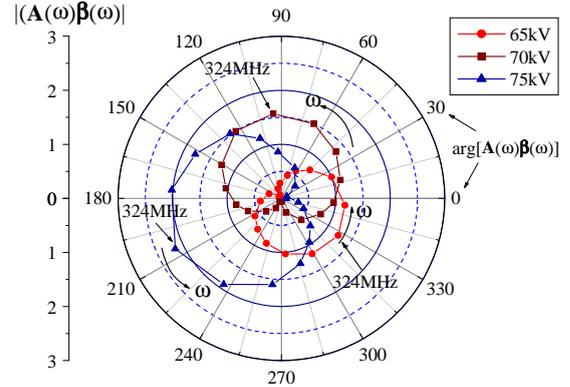

Figure 6: Curves of $A(\omega)\beta(\omega)$ as frequency of 322 ~326MHz under beam voltages of 65kV, 70kV, and 75kV.

## 4 CONCLUSION

A mechanism to produce returning electrons in a collector was successfully simulated using EGS4. A simple analysis was conducted, which took into account the returning electrons, and was in close agreement with the experiment results. Although reversing electrons are sometimes produced at the output cavity and cause instabilities, those were not analyzed in the present paper, which focused on the oscillations due to returning electrons from the collector.